\documentstyle[aps,epsfig]{revtex}
\begin{document}

  \newcommand{\bmit}{$^1$}
  \newcommand{\asu}{$^2$}
  \newcommand{\umass}{$^3$}
  \newcommand{\jgu}{$^4$}
  \newcommand{\fsu}{$^5$}
  \newcommand{\iasa}{$^6$}
  \newcommand{\csu}{$^7$}
  \newcommand{\tu}{$^8$}
  \newcommand{\uutr}{$^{9}$}
  \newcommand{\odu}{$^{10}$}
  \newcommand{\uiuc}{$^{11}$}

\wideabs{
\title{Relativistic effects and two-body currents 
in ${\bf ^2}$H${\bf (\vec{e},e^{\prime}p)n}$ using out-of-plane detection}

\author{
Z.-L.~Zhou,\bmit$^,$\cite{zzhou} $ $ 
J.~Chen,\bmit $ $ 
S.-B.~Soong,\bmit $ $ 
A.~Young,\asu $ $ 
X.~Jiang,\umass$^,$\cite{xdjiang} $ $ 
R.~Alarcon,\asu $ $ 
H.~Arenh\"{o}vel,\jgu $ $
A.~Bernstein,\bmit $ $ 
W.~Bertozzi,\bmit $ $ 
J.~Comfort,\asu $ $ 
G.~Dodson,\bmit$^,$\cite{gdodson} $ $ 
S.~Dolfini,\asu $ $ 
A.~Dooley,\fsu $ $ 
K.~Dow,\bmit $ $ 
M.~Farkhondeh,\bmit $ $ 
S.~Gilad,\bmit $ $ 
R.~Hicks,\umass $ $ 
A.~Hotta,\umass $ $ 
K.~Joo,\bmit$^,$\cite{kjoo} $ $ 
N.I.~Kaloskamis,\iasa $ $ 
A.~Karabarbounis,\iasa $ $ 
S.~Kowalski,\bmit $ $ 
C.~Kunz,\bmit $ $ 
D.J.~Margaziotis,\csu $ $ 
C.~Mertz,\asu $ $ 
M.~Miller,\bmit $ $ 
R.~Miskimen,\umass $ $ 
T.~Miura,\tu $ $ 
H.~Miyase,\tu $ $ 
C.N.~Papanicolas,\iasa $ $ 
G.~Peterson,\umass $ $ 
A.~Ramirez,\asu $ $ 
D.~Rowntree,\bmit $ $ 
A.J.~Sarty,\fsu$^,$\cite{ajsarty} $ $ 
J.~Shaw,\umass $ $ 
T.~Suda,\tu $ $ 
T.~Tamae,\tu $ $ 
D.~Tieger,\bmit $ $ 
J.A.~Tjon,\uutr $ $
C.~Tschalaer,\bmit $ $ 
E.~Tsentalovich,\bmit $ $ 
W.~Turchinetz,\bmit $ $ 
C.E.~Vellidis,\iasa $ $ 
G.A.~Warren,\bmit$^,$\cite{kjoo} $ $ 
L.B.~Weinstein,\odu $ $ 
S.~Williamson,\uiuc $ $ 
J.~Zhao,\bmit $ $ 
and T.~Zwart\bmit  $ $ \\[1.0mm]
(The MIT-Bates OOPS Collaboration)\\[1.0mm]
}

\address{
{\bmit  Laboratory for Nuclear Science, 
        Massachusetts Institute of Technology, Cambridge, MA 02139, USA} \\
{\asu   Department of Physics and Astronomy,
        Arizona State University, Tempe, AZ 85287, USA} \\
{\umass Department of Physics, 
         University of Massachusetts at Amherst, Amherst, MA 01003, USA} \\
{\jgu   Institut f\"ur Kernphysik, Johannes Gutenberg-Universit\"at, 
           D-55099 Mainz, Germany} \\
{\fsu   Department of Physics,
        Florida State University, Tallahassee, FL 32306, USA} \\
{\iasa  Institute of Accelerating Systems and Applications and 
           Department of Physics, University of Athens, Athens, Greece} \\
{\csu   Department of Physics and Astronomy, 
        California State University at Los Angeles, 
        Los Angeles, CA 90032, USA} \\
{\tu    Laboratory of Nuclear Science, 
            Tohoku University, Sendai 982-0826, Japan}\\
{\uutr  Institute for Theoretical Physics, University of Utrecht, 
            3584 CC Utrecht, The Netherlands}\\
{\odu   Department of Physics,
        Old Dominion University, Norfolk, VA 23529, USA} \\
{\uiuc  Nuclear Physics Laboratory, 
        University of Illinois, Urbana, IL 61820, USA} 
}

\draft
\date{\today}
\maketitle

\begin{abstract}
Measurements of the ${^2}$H($\vec{\rm e}$,e$^{\prime}$p)n reaction 
were performed using an 800-MeV polarized electron beam 
at the MIT-Bates Linear Accelerator and with
the out-of-plane magnetic spectrometers (OOPS).
The longitudinal-transverse, 
$f_{\rm LT}$ and $f_{\rm LT}^{\,\prime}$, and the transverse-transverse,
$f_{\rm TT}$, interference responses at a missing momentum of 210 MeV/c
were simultaneously extracted in the dip region at $Q^2=0.15$ (GeV/c)$^2$.
On comparison to models of deuteron electrodisintegration, 
the data clearly reveal strong effects of 
relativity and final-state interactions, 
and the importance of the two-body meson-exchange currents and 
isobar configurations.
We demonstrate that these effects can be disentangled and studied
by extracting the interference response functions using the 
novel out-of-plane technique.
\end{abstract}
}


Careful studies of the deuteron are fundamental to nuclear physics.  
Due to its relatively simple structure, reliable calculations can be 
performed in both non-relativistic and relativistic 
models 
\cite{Arenhoevel97,Mosconi,Tjon,Jeschonnek,Gross,Forest}
for a given nucleon-nucleon (NN) potential,
making the deuteron the first testing ground 
for any realistic nuclear model.  
The electromagnetic probe is of particular importance 
because it is well understood and weak enough  
to allow a simple perturbative interpretation of the observables in terms 
of charge and current matrix elements. 
For these reasons, the electrodisintegration of the deuteron provides
precise information on both the ground-state wave
function \cite{Bernheim,Turk,mainzdeut,zhouprl} 
and the electromagnetic currents 
arising from meson-exchange (MEC) and isobar configurations (IC)
\cite{Arenhoevel82,Gilad}.  
As the deuteron 
is often used as a neutron target,
such detailed understanding of both its structure and 
currents, as well as the dynamics of 
final-state interactions (FSI), is crucial
for applications such as the extraction of precise information on the neutron 
electromagnetic form factors \cite{Bruins,nikhefgen,mainzgen}. 
While in the past the study of realistic 
NN potentials has been the main point of interest, 
the roles of MEC and IC, and the question of relativistic 
corrections (RC), have come into focus recently.

Stringent constraints on nuclear models can be
provided through measurements of the individual interference 
response functions in electron--deuteron 
scattering \cite{Arenhoevel97,Tjon,Jeschonnek}.  
The reason for this is that 
small but dynamically interesting amplitudes 
can be considerably amplified by 
interference with dominant amplitudes and thus may become accessible.  
For example, the longitudinal-transverse response 
$f_{\rm LT}$ is particularly sensitive to the 
inclusion of relativistic effects \cite{Scha92},
while the so-called fifth response $f_{\rm LT}^{\,\prime}$ arises purely through 
final-state interactions \cite{Dolfini}.  
The transverse--transverse response $f_{\rm TT}$ appears to be  mostly
sensitive to MEC and IC \cite{Pellegrino}, 
and this sensitivity increases as the kinematics are moved 
away from the quasielastic (QE) ridge.
By properly choosing kinematical regimes and performing systematic studies 
of these three response functions,
the role played by various interaction effects can 
be disentangled \cite{Gilad}.  

However, very few data on $f_{\rm LT}^{\,\prime}$ and $f_{\rm TT}$ 
exist \cite{Dolfini,Pellegrino,Tamae}. 
This is due to the fact that they
require the detection of protons out of the electron scattering plane
which became possible only recently.
A limited set of data on the cross-section asymmetry, 
$A_{\rm LT}$, or $f_{\rm LT}$ \cite{Scha92,Duc94,Jordan,From94,Bulten,Kasdorp}
is available mainly in QE kinematics and in 
the region of low missing momentum. 
The data were obtained 
with sequential measurements left and right of the momentum transfer, 
which may be vulnerable to systematic errors in 
aspects such as
luminosity variations and kinematic phase-space matching 
when forming $A_{\rm LT}$ or extracting $f_{\rm LT}$. 
Obtaining precise and consistent 
data on all three response functions 
is therefore desirable, in particular in the region of 
high missing momenta where the sensitivity to the various 
currents and dynamical effects is large.
This is precisely the aim of the unique out-of-plane 
spectrometer facility (OOPS) 
at the MIT-Bates Linear Accelerator. 
Recently, we exploited this novel technique
of performing precise extractions of the
interference responses by simultaneous and symmetric
measurements about the direction of the momentum transfer.
This method minimizes possible systematic uncertainties in the extraction.  
Furthermore, we made
simultaneous measurements of these interference response functions
over a wide kinematical region, especially where 
the effects to be studied are enhanced.  

In the one-photon exchange approximation, 
the cross section for the ${^2}$H($\vec{\rm e}$,e$^{\prime}$p)n reaction 
with an unpolarized target 
can be written with five independent terms 
as a function of the energy and momentum transfer ($\omega$, $q$) and 
the polar and azimuthal angles of knocked-out protons with respect to
the momentum transfer direction in the center-of-mass frame of 
the np pair ($\theta_{\rm pq}^{\rm cm}$, $\phi_{\rm pq}^{\rm cm}$)
\cite{Arenhoevel79}: 
\begin{eqnarray} \label{eq: crosssection}
\displaystyle\frac{d^5 \sigma}{d \omega d \Omega_{\rm e} d \Omega_{\rm p}}
    =
c [   \rho_{\rm L}  f_{\rm L} ~+~
      \rho_{\rm T}  f_{\rm T} 
  ~ + ~  
      \rho_{\rm LT}   f_{\rm LT}  \cos (   \phi_{\rm pq}^{\rm cm} ) \nonumber\\
  ~ + ~  
      \rho_{\rm TT}   f_{\rm TT}  \cos ( 2 \phi_{\rm pq}^{\rm cm} )
  ~ + ~  
      h \rho_{\rm LT}^\prime  f_{\rm LT}^\prime  \sin (\phi_{\rm pq}^{\rm cm})].
\end{eqnarray}
Here $c$ is proportional to the Mott cross section,
$h$ is the helicity ($\pm 1$) of the incident electrons, 
$\rho$ are the virtual-photon density matrix elements which depend only
on the electron kinematics,
and $f$ are the response functions 
in the center-of-mass system 
as functions of $\omega$, $q$, and $\theta_{\rm pq}^{\rm cm}$.
In ``parallel" kinematics where
$\theta_{\rm pq}^{\rm cm}$ = 0, the interference 
response functions ($f_{\rm LT}$, $f_{\rm LT}^{\,\prime}$, 
and $f_{\rm TT}$) vanish.
The angle $\theta_{\rm pq}^{\rm cm}$ is directly
related to the missing momentum, $\vec{p}_{\rm m}$, 
which is the difference 
between  $\vec{q}$ and the ejected proton momentum. 
In the plane-wave impulse approximation, $\vec{p}_{\rm m}$ is equal to 
the initial proton momentum in the deuteron.

By measuring the differential cross sections 
at fixed values of $\omega$, $q$, and $\theta_{\rm pq}^{\rm cm}$, but 
at angles $\phi_{\rm pq}^{\rm cm} = 0^{\circ}$, 90$^{\circ}$ 
(and/or 270$^{\circ}$) 
and 180$^{\circ}$ around the $\vec{q}$ vector, 
one can extract the interference responses: 
\begin{eqnarray}
f_{\rm LT} & = & 
      [ d\sigma_{\phi=0^\circ} - d\sigma_{180^\circ} ] 
                 \cdot (2c\rho_{\rm LT})^{-1} \;,  \\\label{eq: flt} 
f_{\rm LT}^{\prime} & = & 
      [ d\sigma_{90^\circ}^{(h=+1)} - d\sigma_{90^\circ}^{(h=-1)} ] 
      \cdot (2c\rho_{\rm LT}^\prime)^{-1}  \;, \\ \label{eq: fltprime} 
f_{\rm TT} & = & 
      [ d\sigma_{0^\circ} + d\sigma_{180^\circ} 
                 - 2 d\sigma_{90^\circ} ] 
                 \cdot (4c\rho_{\rm TT})^{-1} \;, \label{eq: ftt} 
\end{eqnarray}
and form various asymmetries to study the contributions of each individual 
interference term to the cross-section: 
\begin{eqnarray}
A_{\rm LT} & = & 
   \displaystyle\frac { d\sigma_{0^\circ} - d\sigma_{180^\circ} } 
            { d\sigma_{0^\circ} + d\sigma_{180^\circ} } 
                     =  
   \displaystyle\frac{\rho_{\rm LT} f_{\rm LT}}
      { \rho_{\rm L} f_{\rm L} + \rho_{\rm T} f_{\rm T} + 
        \rho_{\rm TT} f_{\rm TT} },  \\\label{eq: alt} \nonumber \\
A_{\rm LT}^{\prime} & = & 
   \displaystyle\frac{ d\sigma_{90^\circ}^{(+1)} - d\sigma_{90^\circ}^{(-1)} } 
   { d\sigma_{90^\circ}^{(+1)} + d\sigma_{90^\circ}^{(-1)} } 
                     =  
   \displaystyle\frac { \rho_{\rm LT}^{\prime} f_{\rm LT}^{\prime} }
      { \rho_{\rm L} f_{\rm L} + \rho_{\rm T} f_{\rm T} - 
        \rho_{\rm TT} f_{\rm TT} }, \\ \label{eq: altprime} \nonumber \\
A_{\rm TT} & = & 
   \displaystyle\frac { d\sigma_{0^\circ} + d\sigma_{180^\circ} - 
                        2d\sigma_{90^\circ} } 
            { d\sigma_{0^\circ} + d\sigma_{180^\circ} + 2d\sigma_{90^\circ} } 
                     =  
   \displaystyle\frac{\rho_{\rm TT} f_{\rm TT}}
      { \rho_{\rm L} f_{\rm L} + \rho_{\rm T} f_{\rm T} } \; . \label{eq: att} 
\end{eqnarray}


The experiment was carried out with the OOPS system 
\cite{DolfiniNIM,Mandeville} used to detect knock-out 
protons in coincidence with electrons detected in the reconfigured 
one-hundred-inch proton spectrometer (OHIPS) \cite{xjiangthesis}.
We developed a detailed computer-aided alignment method which 
ensured a precise alignment of the OOPS with
absolute accuracies in position and angles 
better than 0.3 mm and 0.3 mrad, respectively \cite{joecomfort}.
Spectrometer optics data were obtained for all OOPS modules and OHIPS.
We performed ${^1}$H(e,e$^{\prime}$p) coincidence 
studies between OHIPS and the OOPS,
and extracted absolute cross sections which were within 
2\% of the expected values \cite{sbsoongthesis,jchenthesis}. 
Calculations of the coincidence phase-space, the effect of the 
radiative tail and multiple scattering
corrections were carried out by Monte-Carlo simulations \cite{vellidis}. 

The measurements of the ${^2}$H($\vec{\rm e}$,e$^{\prime}$p)n reaction 
were performed by using an 800-MeV, 1\% duty-factor polarized electron beam 
with an average current of 4 $\mu$A and 
a 160-mg/cm$^2$ thick liquid deuterium target. 
The polarization of the electron beam was measured with 
the B-line {M\o ller} polarimeter
and averaged to be $38.6\pm 4.0$\%. 
The OHIPS was positioned at a scattering angle of $\theta_{\rm e}=31.0^{\circ}$
and its central momentum was set to 645.0 MeV/c, corresponding to  
$q = 414$ MeV/c, $\omega = 155$ MeV and $x_{\rm Bjorken} = 0.52$. 
Three OOPS were positioned 
at $\phi_{\rm pq}^{\rm cm} = 0^{\circ}$, 90$^{\circ}$ and 180$^{\circ}$ 
with $\theta_{\rm pq}^{\rm cm}$ fixed at 38.5$^{\circ}$,
thus providing simultaneous measurements of
all three interference responses 
$f_{\rm LT}$, $f_{\rm LT}^{\,\prime}$, and $f_{\rm TT}$ as well as their associated asymmetries
at $p_{\rm m}=210$ MeV/c \cite{jchenthesis}. 
In these kinematics the signal-to-noise ratio
in the most forward OOPS was about 1:1. 
Here, we report on the results from the measurements in the ``dip" region
between the QE ridge ($\omega\simeq 90$ MeV, $x_{\rm Bjorken}\simeq 1$) 
and the $\Delta$ resonance. 
Measurements of the $f_{\rm LT}$ and $f_{\rm LT}^{\,\prime}$ response functions
on top of the QE ridge 
were also performed \cite{zhoublast} and will be reported on later.

\begin{figure}[htb]\unitlength1cm
\begin{picture}({8},{6.4})
\put(+0.7,-1.5){\epsfxsize=6.5cm \epsfbox{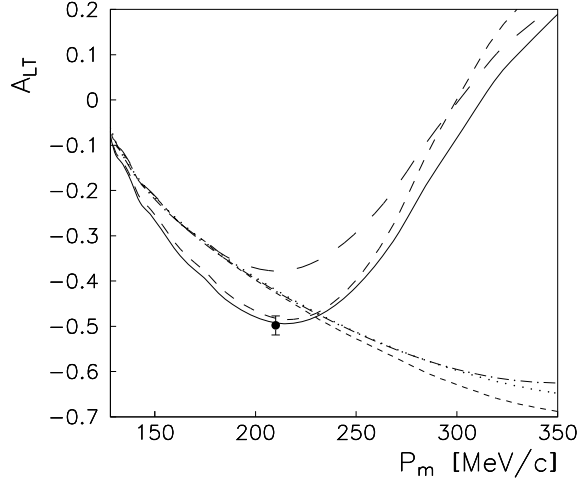}}
\end{picture}
\caption{
The asymmetry $A_{\rm LT}$ as a function of $p_{\rm m}$.
The calculations by Arenh\"ovel {\sl et al.} [1]: 
PWBA+RC (short-dashed), FSI+RC (dashed), FSI+MEC+IC (long-dashed)
and full -- FSI+MEC+IC+RC (solid). 
Tjon {\sl et al.} [3] (PWBA):
dash-dotted curve;
while $\sigma_{\rm cc1}$ [34]:
the dotted curve.
}
\label{alt} \end{figure}

In order to suppress possible systematic uncertainties in 
extracting the data and to reduce the effects of
kinematic broadening when comparing to calculations,
only events from the overlapping portion of
the detector acceptances are selected 
after matching the $\phi_{\rm pq}^{\rm cm} = 0^\circ$ and $180^\circ$
detector phase-spaces. 
Fig. \ref{alt} shows the asymmetry $A_{\rm LT}$ 
as a function of $p_{\rm m}$. 
The PWBA(plane wave Born approximation)+RC 
result of Arenh\"ovel {\sl et al.} \cite{Arenhoevel97} 
is compared to 
the full relativistic calculations (PWBA) by Tjon {\sl et al.} \cite{Tjon} 
and the $\sigma_{\rm cc1}$ prescription of de Forest \cite{deForest} 
for the off-shell electron-proton cross section.
It seems that all these plane-wave approaches 
do not differ much and they obviously fail to describe the data. 

The results show that 
the asymmetry is strongly sensitive to final-state interactions.
The calculations with simple plane-wave approximations 
are inadequate and consequently a rigorous effort 
in including FSI is needed. 
Also, relativistic current operators or relativistic corrections are necessary.
The Bonn potential of the NN interactions \cite{Bonn} 
is used in the calculations of Arenh\"ovel {\sl et al.},
but the predictions show little sensitivity to the choice of 
other realistic NN potentials for these kinematics \cite{Arenhoevel97}.
In addition, the calculations show very little sensitivity to 
the two-body currents for $A_{\rm LT}$.

\begin{figure}[htb]\unitlength1cm
\begin{picture}({8},{5.7})
\put(+0.7,-1.7){\epsfxsize=6.5cm \epsfbox{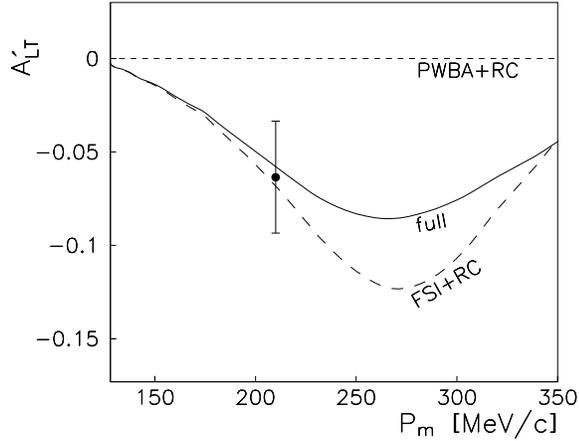}}
\end{picture}
\caption{
The asymmetry $A_{\rm LT}^{\,\prime}$ as a function of $p_{\rm m}$.
Calculations by Arenh\"ovel {\sl et al.} with ingredients as labeled.
}
\label{altprime} \end{figure}

The out-of-plane detection also makes it possible to measure
the helicity asymmetries $A_{\rm LT}^{\,\prime}$, which represent the
 imaginary part of the interference of the longitudinal and transverse 
current matrix elements.
Fig. \ref{altprime} shows $A_{\rm LT}^{\,\prime}$ 
for the OOPS at $\phi_{\rm pq}^{\rm cm} = 90^\circ$.
The calculations by Arenh\"ovel {\sl et al.} 
agree well with the data when the FSI are included,
because $A_{\rm LT}^{\,\prime}$ 
arises entirely from complex amplitudes interfering in 
the final-state processes.
In PWBA, where only real amplitudes are involved,
the asymmetry vanishes.
In addition, as indicated by Eq. \ref{eq: crosssection}, 
when $\phi_{\rm pq}^{\rm cm}=0^\circ$ or $180^\circ$, $A_{\rm LT}^{\,\prime}$ vanishes.
As a consistency check, our data in the two in-plane OOPS together 
yielded an asymmetry of $0.006 \pm 0.009$. 

It is interesting to observe that 
the asymmetries are opposite in sign
to the $p$-shell proton knock-out of 
the $^{12}$C($\vec{\rm e}$,e$^{\prime}$p)$^{11}$B 
reaction \cite{xjiangthesis,mandprl},
as shown by both data and calculations.
In the low missing momentum region
of the deuteron electrodisintegration, 
the FSI is dominated by the spin--spin interactions of the np pair,
while in the p-$^{11}$B interactions,
the spin--orbit parts are more important.

\begin{figure}[htb]\unitlength1cm
\begin{picture}({8},{6.5})
\put(+0.7,-1.5){\epsfxsize=6.5cm \epsfbox{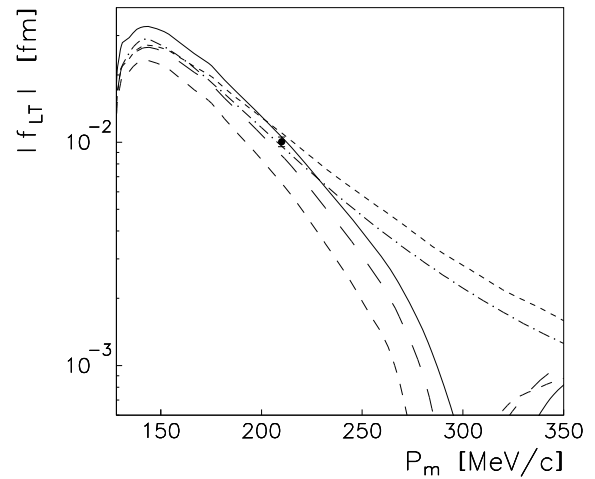}}
\end{picture}
\caption{
The response function $f_{\rm LT}$
as a function of 
$p_{\rm m}$.
The calculations by Arenh\"ovel {\sl et al.}:
PWBA (short-dashed), FSI (dashed), FSI+MEC+IC (long-dashed), 
and full (solid), while by Tjon {\sl et al.}: PWBA+FSI (dash-dotted).
}
\label{dip_flt} \end{figure}

The determination of the absolute cross section makes it possible 
to extract also individual response functions.
In Asymmetries, the denominators 
may cancel some of the effects in the numerators, 
as can be seen in our $f_{\rm LT}$ data shown in  Fig. \ref{dip_flt}.
Here, the data are again compared to calculations 
by Arenh\"ovel {\sl et al.}
Not only relativistic corrections 
and  detailed calculations of FSI, 
but also the two-body currents, 
are needed in order to bring the predictions
into agreement with the data. 
However, one observes substantial cancellations 
between the effects of two-body currents and FSI.
Accordingly,
the calculations by Tjon {\sl et al.}, which include 
only limited contributions from FSI, but do not contain 
two-body currents, are also in agreement with the data.

\begin{figure}[htb]\unitlength1cm
\begin{picture}({8},{6.5})
\put(+0.7,-1.5){\epsfxsize=6.5cm \epsfbox{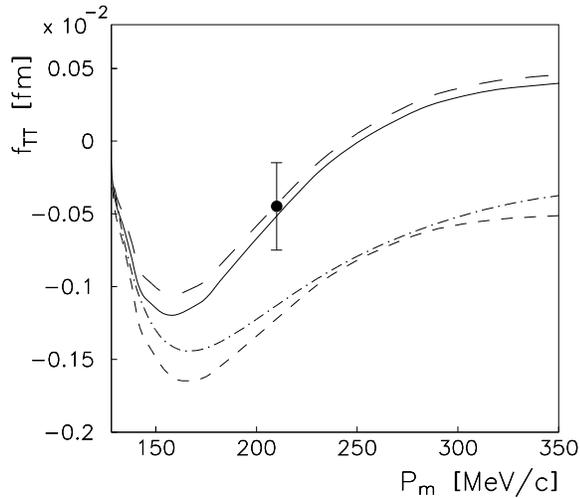}}
\end{picture}
\caption{
The response function $f_{\rm TT}$
as a function of $p_{\rm m}$.
The calculations by Arenh\"ovel {\sl et al.}:
FSI+RC (dashed), FSI+MEC+IC (long-dashed)
and full (solid), and 
by Tjon {\sl et al.}: PWBA+FSI (dashed-dotted).
}
\label{dip_ftt} \end{figure}

Isolating the contributions of the 
two-body currents from other competing reaction effects
can be done by separating the remaining interference response, $f_{\rm TT}$.
As shown in Fig. \ref{dip_ftt},
various models predict that 
$f_{\rm TT}$ (or $A_{\rm TT}$) is strongly sensitive to 
the two-body currents while they do not depend so much on
the relativistic effects, in contrast to $A_{\rm LT}$ and $f_{\rm LT}$.
Our data agree with the full calculations 
by Arenh\"ovel {\sl et al.} \cite{Arenhoevel97}
which have recently been improved by including retardation diagrams. 
The calculations by Tjon {\sl et al.} which currently do not contain 
two-body contributions, fail to describe the data.
The $f_{\rm TT}$ data 
demonstrate the power of using out-of-plane detection
and new observables to isolate such small, but interesting contributions 
to the electromagnetic currents in deuteron disintegration.

In summary, our data 
clearly reveal strong effects of
relativity and FSI, as well as of two-body
currents arising from MEC and IC.
We conclude that the two-body currents and relativity are extremely 
important to the understanding of the deuteron.
In order to describe the data better, more rigorous relativistic 
calculations including all ingredients discussed here are needed.
We show that competing effects in 
the deuteron electrodisintegration can be probed selectively
by studies of multiple interference response functions 
using the novel out-of plane technique.

Data with higher statistical precision 
will be taken in the near future, especially 
in the region of higher missing momentum ($> 250$ MeV/c)
and as a function of the energy transfer 
up to the $\Delta$ resonance \cite{zhou1997}.
This comprehensive set of data will clarify 
the role of relativity and two-body currents, and also 
provide a detailed understanding of the isobar configurations
and possible knowledge of the $\Delta$--$N$ interactions.
The figure of merit in measuring $A_{\rm LT}^{\,\prime}$ will be 
improved by an order of magnitude
when a highly polarized continuous-wave beam 
is used and two OOPS
are mounted simultaneously at $\phi_{\rm pq}^{\rm cm} = \pm 90 ^\circ$, 
which is now possible.  

We would like to thank the Bates staff in making this experiment possible.
This work is supported in part by the US Department of Energy and the National
Science Foundation, Grant-in-Aid for International Scientific Research
by the Ministry of Education, Science, and Culture in Japan, and
Deutsche Forschungsgemeinschaft.


\end{document}